\documentclass[12pt]{article}

\usepackage{amssymb,amsmath,fullpage}
\usepackage{amsfonts}
\usepackage{graphicx}
\usepackage{textcomp,gensymb} 

\newcommand{\ff}{\mathbf{f}}
\newcommand{\MM}{\mathbf{M}}
\newcommand{\uu}{\mathbf{u}}
\newcommand{\PP}{\mathbf{P}}
\newcommand{\nn}{\mathbf{n}}
\newcommand{\pp}{\mathbf{p}}
\newcommand{\bphi}{\mathbf{\Phi}}

\newcommand{\ds}{\displaystyle}
\newcommand{\be}{\begin{eqnarray}}
\newcommand{\ee}{\end{eqnarray}}

\newcommand{\ie}{{\it i.e.}\ }
\newcommand{\eg}{{\it e.g.}\ }
\newcommand{\etal}{{\it et al.}\ }

\newcommand{\nt}{\bar{n}}
\newcommand{\pt}{\bar{p}}
\newcommand{\jn}{{j^n}}
\newcommand{\jp}{{j^p}}
\newcommand{\fp}{{{F}^P}}

\begin{document}

\title{A fast and robust numerical scheme for solving models of charge carrier transport and ion vacancy motion in perovskite solar cells}

\author{N.E. Courtier\thanks{Mathematical Sciences, University of Southampton, SO17 1BJ, UK} , G. Richardson$^{*}$, \& J.M. Foster\thanks{Department of Mathematics, University of Portsmouth, PO1 2UP, UK}}

\maketitle

\begin{abstract}
\noindent
Drift-diffusion models that account for the motion of both electronic and ionic charges are important tools for explaining the hysteretic behaviour and guiding the development of metal halide perovskite solar cells. Furnishing numerical solutions to such models for realistic operating conditions is challenging owing to the extreme values of some of the parameters. In particular, those characterising (i) the short Debye lengths (giving rise to rapid changes in the solutions across narrow layers), (ii) the relatively large potential differences across devices and (iii) the disparity in timescales between the motion of the electronic and ionic species give rise to significant \emph{stiffness}. We present a finite difference scheme with an adaptive time step that is posed on a non-uniform staggered grid that provides second order accuracy in the mesh spacing. The method is able to cope with the stiffness of the system for realistic parameters values whilst providing high accuracy and maintaining modest computational costs. For example, a transient sweep of a current-voltage curve can be computed in only a few minutes on a standard desktop computer.
\end{abstract}


\section{Introduction}
Recent rapid improvements in power conversion efficiency have brought metal halide perovskite solar cells (PSCs) to the forefront of the emerging thin-film photovoltaic technologies. Efficiencies in excess of 20\% \cite{Correa-Baena2017,Stolterfoht2017}, comparable to that of standard crystalline silicon devices, have been achieved using methylammonium lead tri-iodide and other perovskite materials as absorber layers in thin film architectures \cite{Kim2012,Lee2012}. This high performance, together with their relatively low cost of manufacture, mean that the continued development of PSCs is an extremely active area of research. Recent reviews of the field have been given in \cite{Miyasaka2015,Niu2015,Park2013,Stranks2015,Sum2015}.

Planar PSC architectures are formed by a perovskite absorber layer sandwiched between an electron transporting layer (ETL) and a hole transport layer (HTL). The most commonly used perovskite absorbing material is methylammonium lead tri-iodide (CH$_3$NH$_3$PbI$_3$) \cite{Jacobsson2016}, however, recently other mixed halide formulations (in which some of the iodide ions are substituted by other halides) and mixed cation formulations (in which the methylammonium cation is fully or partially replaced with formamidinium and/or caesium) have also been successfully employed \cite{Pellet2014,Saliba2016,Stolterfoht2017}. Numerous different materials have been used as ETLs and HTLs but common choices are titania (TiO$_2$) for the former and spiro (2,2',7,7'-tetrakis-(N,N-di-p-methoxyphenylamine)-9,9'-spirobifluorene) for the latter. 

\begin{figure}
\centering
\scalebox{1.25}{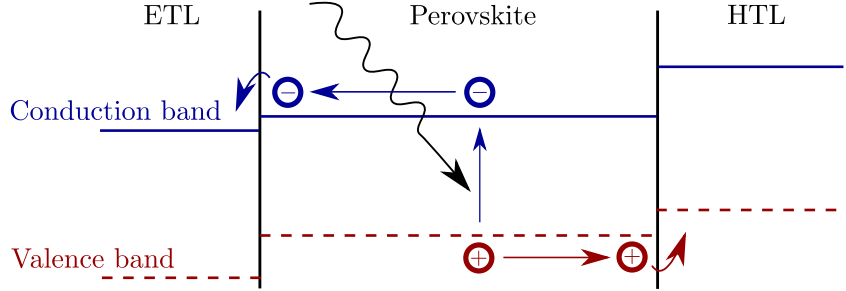}
\caption{Sketch of a planar PSC showing (i) the structure of the conduction (blue solid lines) and valence (red dashed lines) bands and (ii) the flow of electrons and holes.}
\label{schematic}
\end{figure}

Under illumination, incident photons with energies above the band gap are absorbed in the perovskite absorbing layer generating weakly bound excitons (binding energy $\sim$50meV \cite{Koutselas1996}). These excitons readily dissociate into a free electron in the conduction band and a hole in the valence band which can move independently under the influences of both thermally-induced diffusion and electronically-induced drift. Since (i) the conduction band in the HTL is above that in the perovskite and (ii) the valence band in the ETL is below that in the perovskite, barriers exist for the entry of holes and electrons into the ETL and HTL respectively. The differences in the band energies of the different semiconductors also lead to the formation of a built-in field in the perovskite which encourages electrons to move towards the ETL and holes towards the HTL, thereby setting up a useful electronic current. A cartoon of this process is shown in figure \ref{schematic}.

One notable peculiarity of PSCs is their long timescale transient behaviour occurring on the order of 10s of seconds. This is observed in (i) current-voltage (J-V) curve `hysteresis' \cite{Snaith2014,Unger2014}
and (ii) current transients \cite{OKane2017,Unger2014}.
Initially, three possible explanations for these slow dynamics were postulated \cite{Snaith2014}, namely: (a) the formation of ferroelectric domains (b) large-scale trapping of electrons, and (c) mobile ions. However, as discussed in \cite{Richardson2016}, there is a growing consensus that the only mechanism that offers a coherent explanation of the observed data is the slow motion of positively charged anion vacancies. More recently, very long timescale reversible transients in the efficiency of PSCs have been observed over periods of several hours \cite{Domanski2017}. These have been attributed to the motion of slow cation vacancies.

A range of approaches exist to modelling PSC behaviour that extend from fundamental atomistic density functional theory (DFT) calculations (\eg \cite{Eames2015}) to equivalent circuit device models (\eg \cite{Cojocaru2015}). High computational costs mean that DFT can only be applied to a few atoms and over extremely short timescales. At the other extreme, equivalent circuit models, which are very straightforward to solve, are hard to connect directly to the device physics. An intermediate path, which leads to a computationally tractable model that can be directly interpreted in terms of device physics, is given by charge transport, or equivalently drift-diffusion, modelling. Notably this approach allows (as in \cite{Richardson2016}) for parameters obtained from DFT calculations on the perovskite structure to be incorporated into the model. It has also been used in a variety of other solar cell applications (for example in organics \cite{Buxton2006,Foster2013,Richardson2017}). Initially the importance of ion motion in PSCs was not fully appreciated and as a consequence charge transport models, that treated only electron and hole transport, were formulated \cite{Foster2014,Li2013}. Subsequently models that considered the coupling between ion vacancy motion and charge transport were investigated in \cite{Calado2016,Domanski2017,Gottesman2016,Jacobs2017,OKane2017,Richardson2016,VanReenen2015}. However the incorporation of realistic (high) densities of ion vacancies leads to a model that is computationally challenging to solve owing to (i) narrow Debye layers across which rapid changes in solution occur, (ii) very large changes in the magnitude of  the solution across the device and (iii) the large disparity between the timescales for ion vacancy motion and electron and hole transport. Of the aforementioned works on charge transport modelling of PSCs, only \cite{Courtier2017,OKane2017,Richardson2016} manage to obtain solutions in physically relevant parameter regimes. However in all these studies the analysis relies on making some level of asymptotic approximation to the governing equations. Whilst these approximations are well-justified in most scenarios, and lead to accurate approximate solutions, a numerical treatment of the full system of equations, that is capable of furnishing solutions across all relevant timescales and operating regimes without simplification, is highly desirable. This motivates the aim of this work which is to present a numerical method capable of accurately solving such problems using modest computational power.

The work is set out as follows. In the next section, we outline the charge transport model in dimensionless form, describe how the system is related to its dimensional counterpart and give estimates for the sizes of the dimensionless parameters. In \S\ref{scheme} we describe the spatial (finite difference) discretisation. Then, in \S\ref{implementation}, we outline implementation of the method in MATLAB \cite{MATLAB} using the Advanpix Multi-Precision Toolbox \cite{mct17}. In \S\ref{verify} we benchmark the scheme using some experimentally motivated test cases, comparing the accuracy of the numerical approach to the asymptotic approach developed in \cite{Courtier2017,Richardson2009} and demonstrating the pointwise convergence of the method. In \S\ref{discussion}, we briefly discuss how this approach compares to other methods that we tried, including spectral methods and MATLAB's built-in solver PDEPE. Finally, in \S\ref{conclusions}, we present our conclusions.


\section{\label{equations}The model}

Here we follow \cite{Richardson2016} and consider a model that has been formulated for charge transport in a planar PSC consisting of a perovskite absorber layer sandwiched between highly doped electron and hole transport layers. The doping, and the resulting high conductivity of the transport layers, allows us to treat these layers as 'quasi-metals' through which the electric potential is uniform and equal to that on their respective contacts. In turn this leads to a {\it single-layer model} in which all the relevant physics takes place within the perovskite layer. The key physical processes described by the model are the motion, generation and recombination of highly mobile charge carriers (electrons and holes) and their interaction with (much less mobile) anion vacancies and a uniformly distributed stationary cation vacancy distribution.

Since the focus of this work is on the numerical method required to solve this model, and detailed descriptions are given in \cite{Richardson2016,Courtier2017}, the model is only outlined here. Notably, the numerical scheme that we develop for the single-layer model can easily be extended to more realistic multi-layer models. For example, a three-layer that incorporates charge carrier dynamics in the electron and hole transport layers. An investigation of this three-layer model will be the subject of future work. 

\subsection{Model equations} 
In order to present the method in the simplest fashion possible, whilst highlighting potential numerical difficulties, the model equations are presented in dimensionless form (for full details of the non-dimensionalisation see \cite{Courtier2017}). The key variables in the problem are time, $t$; the perpendicular distance from the interface of the perovskite layer with the ETL, $x$; the mobile anion vacancy density, $P$; the free-electron density, $n$; the hole density, $p$; the electric field (in the $x$-direction), $E$; the electric potential, $\phi$; the anion vacancy flux (in the $x$-direction), $\fp$; and the electron and hole current densities, $\jn$ and $\jp$, respectively. 
The (dimensionless) model consists of three conservation equations for the three mobile charged species:
\be
\label{Peqn} \frac{\partial P}{\partial t} +\lambda\frac{\partial \fp}{\partial x}=0,
\quad \mbox{where} \quad \fp = -\left(\frac{\partial P}{\partial x}-PE\right), \\
\label{neqn} \nu \frac{\partial n}{\partial t} = \frac{\partial \jn}{\partial x}+G(x)-R(n,p),
\quad \mbox{where} \quad \jn = \kappa_n\left(\frac{\partial n}{\partial x}+nE\right), \\
\label{peqn} \nu \frac{\partial p}{\partial t} = -\frac{\partial \jp}{\partial x}+G(x)-R(n,p),
\quad \mbox{where} \quad \jp = -\kappa_p\left(\frac{\partial p}{\partial x}-pE\right) \label{jp},
\ee
in which $G(x)$ and $R(n,p)$ represent rates of charge carrier generation and recombination, respectively. These conservation equations 
couple to Poisson's equation:
\be
\label{phieqn} \frac{\partial E}{\partial x} = \frac{1}{\lambda^2} \left(P-1+\delta\left(p-n\right) \right)
\quad \mbox{where} \quad E = -\frac{\partial \phi}{\partial x}.
\ee
The term $(P-1+\delta\left(p-n\right))$, in the equation above, is the dimensionless charge density and comprises contributions from anion vacancies, stationary cation vacancies, holes and electrons, respectively.

Equations (\ref{Peqn})-(\ref{phieqn}) hold within the perovskite layer which is bounded by its interface with the ETL, on $x=0$, and its interface with the HTL, on $x=1$. In practice, the three dimensionless parameters $\lambda$ (the ratio of the Debye length $L_d$ to the the perovskite layer width), $\nu$ (the ratio of the timescales for electron and hole motion to that for ion vacancy motion) and $\delta$ (the ratio of typical carrier concentration to typical vacancy concentration) are all very small, see (\ref{stack}). 

\subsection{Boundary and initial conditions}
At the interfaces with the ETL (on $x=0$) and the HTL (on $x=1$) we require that there is no flux of anion vacancies and that the potential is specified (and equal to that in the adjacent contact). In addition $\bar{n}$, the free electron concentration on $x=0$, is determined by the band energy offsets with the ETL; while $\bar{p}$, the hole concentration on $x=1$, is determined by the band energy offsets with the HTL; and $j_p|_{x=0}$ and $j_n|_{x=1}$ are determined by $R_l$ and $R_r$, the rates of surface recombination on these interfaces. It follows that the boundary conditions read:
\be 
\label{leftbcs}
\fp|_{x=0} = 0, \quad \phi|_{x=0} = \frac{\Phi(t)-\Phi_{bi}}{2}, \quad n|_{x=0} = \nt, \quad j_p|_{x=0} = -R_l(p), \\
 \label{rightbcs}
\fp|_{x=1} = 0, \quad \phi|_{x=1} = -\frac{\Phi(t)-\Phi_{bi}}{2}, \quad p|_{x=1} = \pt, \quad j_n|_{x=1} = -R_r(n).
\ee
Here the total potential drop across the cell, $\Phi_{bi}-\Phi(t)$, is split into two parts; a built-in voltage, $\Phi_{bi}$, that arises from the difference in band energies between the ETL and the HTL and an applied voltage, $\Phi(t)$. In this formulation of the problem, potentials are measured in units of the thermal voltage (approximately $0.025$V). The built-in voltage is typically around $1.1$V and standard experiments vary the applied voltage within the range of $-0.5$ to $2$V, therefore the total potential drop across the cell, $\Phi_{bi}-\Phi(t)$, can be quite large. In turn this can lead to very large variations in both ion vacancy and carrier concentrations across the perovskite layer which renders furnishing numerical solutions challenging.

The problem is closed by the following initial conditions for the electron, hole and anion vacancy concentrations:
\begin{equation} \label{ics}
n|_{t=0} = n_{\text{init}}(x), \quad p|_{t=0} = p_{\text{init}}(x), \quad P|_{t=0} = P_{\text{init}}(x).
\end{equation}

\subsection{Definition of dimensionless variables and parameters}
The dimensional counterparts of the variables used in the model (\ref{Peqn})-(\ref{ics}), which we denote by a star,  are retrieved from the following rescaling (see  \cite{Courtier2017}):
\be
\label{nondim}
\begin{array}{cccc}
\ds x^* = b x, &\ds t^* = \frac{L_d b}{D_+} t, &\ds \Phi^* = V_T\Phi, &\ds \Phi_{bi}^* =V_T \Phi_{bi}, \\*[3mm]
\ds p^* = \frac{F_{ph} b}{\hat{D}} p, &\ds n^* = \frac{F_{ph} b}{\hat{D}} n, &\ds P^* = N_0 P, &\ds \jp^* = q F_{ph} {\jp}, \\*[3mm]
\ds \jn^* = q F_{ph}{\jn}, &\ds \fp^* = \frac{D_+ N_0}{b} \fp, & \ds G^* = \frac{F_{ph}}{b} G, &\ds R^* = \frac{F_{ph}}{b} R,\\*[3mm]
\ds  R_l^*=F_{ph} R_l, & \ds R_r^*=F_{ph} R_r.
\end{array}
\ee
Here $b$ denotes the width of the perovskite layer, $L_d$ the Debye length, $D_+$ the anion vacancy diffusion coefficient, $q$ the elementary charge,  
$V_T$ the thermal voltage (\ie $k_B T/q$ where $T$ the absolute temperature), $F_{ph}$ the incident photon flux, $\hat{D}$ a typical electronic charge carrier diffusivity and $N_0$ the equilibrium anion vacancy density. The Debye length is calculated based on the most populous charged species, which in this instance is the anion vacancies, as follows:
\be
L_d = \left( \frac{\varepsilon k T}{q^2 N_0} \right)^{1/2},
\ee
where $\varepsilon$ is the permittivity of the perovskite. The dimensionless parameters in  (\ref{Peqn})-(\ref{ics}) are given in terms of physical constants by the relations:
\be
\label{parameters}
\begin{array}{ccccc}
\ds \nu = \frac{D_+ b}{\hat{D} L_d}, &\ds \kappa_p = \frac{D_p}{\hat{D}}, &\ds  \kappa_n = \frac{D_n}{\hat{D}}, &\ds \lambda = \frac{L_d}{b}, &\ds  \\*[3mm]
\ds \delta = \frac{F_{ph} b}{\hat{D} N_0}, &\ds \nt = \frac{n_b \hat{D}}{F_{ph} b}, &\ds \pt = \frac{p_b \hat{D}}{F_{ph} b}, & \ds \Phi_{bi}=\frac{V_{bi}}{V_T}.
\end{array}
\ee
Here $D_n$ and $D_p$ are free-electron and hole diffusivities, respectively, and $n_b$ and $p_b$ are the (dimensional) electron density on the ETL interface and the (dimensional) hole density on the HTL interface, respectively. 

In \cite{Courtier2017} it is shown that, for a typical planar device formed by a methylammonium lead tri-iodide perovskite absorber layer sandwiched between a titania ETL layer and a spiro HTL layer, the dimensionless parameters defined in (\ref{parameters}) take the values
\be \label{stack}
\begin{array}{cccc}
\ds \nu \approx 5.8 \times 10^{-10}, & \ds \kappa_p \approx 1, &\ds \kappa_n \approx1, &\ds \lambda \approx 2.4 \times 10^{-3},\\*[3mm]
\ds \delta \approx 2.1 \times 10^{-7}, &\ds \nt \approx 20, &\ds \pt \approx 0.30 , &\ds \Phi_{bi} \approx 40.
\end{array}
\ee
The difficulty in solving (\ref{Peqn})-(\ref{jp}) arises from the extreme values of the parameters $\nu$, $\lambda$ and $\Phi_{bi}-\Phi(t)$. The very small value of $\nu$ reflects the large disparity in timescales for the electronic (fast) and ionic (slow) motion. This feature of the problem necessitates the use of an adaptive timestep since any fixed time stepping method would either be prohibitively slow or be incapable of capturing the fast electronic dynamics. As is typical for many electrochemical problems (see \eg \cite{George2015}), the parameter characterising the relative size of the Debye length, in this case $\lambda$, is also extremely small. This gives rise to appreciable stiffness in the system owing to the rapid changes in the solution across the narrow Debye layers. This issue is further exacerbated by the relatively large value of $\Phi_{bi}-\Phi(t)$. The concentrations of the different charge species in the slender Debye layers are approximately Boltzmann distributed, i.e. they vary exponentially with the potential. Thus, a change in the potential of size $20$ (the size of the drop expected across one Debye layer when the drop across the whole device is $40$, e.g., near short-circuit) gives rise to a change in $n$, $p$ or $P$ by a factor of $\exp(20) = O(10^{9})$ across a region of width $O(10^{-3})$. Therefore, what at first glance appears to be merely a stiff problem is actually an extremely stiff problem. Notably,these very large changes in the magnitudes of the  concentrations of the charged species can give rise to large condition numbers and, depending on the computational precision being used, significant round off errors.

\section{\label{scheme}Numerical scheme}
The central technique underlying our numerical scheme is the method of lines. The spatial derivatives in (\ref{Peqn})-(\ref{ics}) are discretised using second order accurate finite differences on a `staggered grid'. Since the equation governing the electric potential is elliptic, in contrast to those for the charge carrier densities which are parabolic, the discrete system takes the form of a system of coupled differential-algebraic equations (DAEs). As such, it requires a specialised algorithm for temporal integration. Here, we employ MATLAB's integrator {\tt ode15s} \cite{MATLAB} to evolve in time. In the physically relevant parameter regimes, typical solutions exhibit rapid changes in the narrow Debye (boundary) layers which gives rise to significant \emph{stiffness} in the DAE system. This difficulty is overcome by (i) employing a non-uniform grid spacing, and (ii) using Advanpix's Multiprecision Computing Toolbox \cite{mct17} to overload MATLAB's native commands with arbitrary-precision counterparts, thereby retaining good accuracy despite the large condition numbers of the underlying matrices. We note that the treatment of the conservation equations used here is in contrast to the Scharfetter-Gummel scheme that is widely used in electrochemical problems \cite{Scharfetter1969}. This difference arises because their discretisation is aimed at addressing issues of transport within drift-diffusion equations whereas our is dealing with the difficulties associated with stiffness.

\subsection{Computational grid}
\begin{figure}
\centering
\scalebox{1}{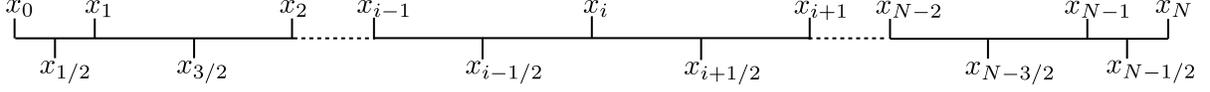}
\caption{A schematic of the computational grid.}
\label{griddiag}
\end{figure}

The computational grid is comprised of $N+1$ arbitrarily positioned grid points, denoted by $x=x_i$ for $i=0,...,N$, which partition the domain $x\in[0,1]$ into $N$ subintervals. We further introduce a set of $N$ `half-points' denoted by $x=x_{i+1/2}$ and defined as follows
\begin{equation}
x_{i+1/2} = \frac{x_{i+1}+x_i}{2} \quad \text{for} \quad i=0,...,N-1.
\end{equation}
A sketch of the grid is shown in figure \ref{griddiag}. The anion vacancy profiles are determined subject to the Neumann conditions (\ref{leftbcs}a) and (\ref{rightbcs}a) which require zero flux of ions at each boundary. Thus, it is natural to compute the ion vacancy flux at the grid points and the ion vacancy density at the half points. Not only does this allow (\ref{leftbcs}a) and (\ref{rightbcs}a) to be imposed straightforwardly, it also ensures that the property of global conservation of anion vacancies is inherited by the discrete system\footnote{Global conservation of anion vacancy concentrations in the discrete system, up to second order, is reflected in the property that $d/dt \left( \sum_{i=0}^{N-1} P_{i+1/2} \right)=0$, see equation (\ref{Pdisc}).}. The equation determining the electric potential is to be solved subject to the Dirichlet conditions (\ref{leftbcs}b) and (\ref{rightbcs}b) which motivates tracking the potential at the grid points and the electric field at the half points. Since the governing equations for electrons (holes) are to be solved subject to one Dirichlet and one Neumann condition, namely (\ref{leftbcs}c) and (\ref{rightbcs}d) ((\ref{leftbcs}d) and (\ref{rightbcs}c)), it is not so clear whether the concentration or flux should be defined at the grid points. Here, we elect to track the concentration at the grid points and the fluxes on the half-points so that evaluating the electron and hole contributions to the charge density on the right-hand side of (\ref{phieqn}a) can be done so without interpolation, thereby avoiding additional errors. In summary we introduce discretised variables defined by
\begin{align}
&P_{i+1/2}=P|_{x=x_{i+1/2}}, &&\fp_{i}=\fp|_{x=x_i},\\
&\phi_i =\phi|_{x=x_i}, &&E_{i+1/2} =E|_{x=x_{i+1/2}},\\
&n_i=n|_{x=x_i}, &&\jn_{i+1/2}=\jn|_{x=x_{i+1/2}},\\
&p_i=p|_{x=x_i},  &&\jp_{i+1/2}=\jp|_{x=x_{i+1/2}}.
\end{align}  

\subsection{Spatial discretisation}
Spatial discretisation will be carried out by introducing the discrete operators $\mathfrak{D}_i$ and $\mathfrak{D}_{i+1/2}$ which approximate the first derivative operator at $x_i$ and $x_{i+1/2}$, respectively, and the discrete averaging/interpolation operators, $\mathfrak{I}_i$ and $\mathfrak{I}_{i+1/2}$ which approximate half points quantities on the grid points and grid point quantities at the half-points, respectively. For a generic variable $w$, these four operators are defined as follows:
\begin{eqnarray}
\label{op1} \frac{\partial w}{\partial x}\Big{|}_{x=x_{i+1/2}} \approx &&\mathfrak{D}_{i+1/2}(w_{i+1},w_i) = \frac{w_{i+1}-w_i}{x_{i+1}-x_{i}},\\
\frac{\partial w}{\partial x}\Big{|}_{x=x_{i}} \approx &&\mathfrak{D}_i(w_{i+1/2},w_{i-1/2}) = \frac{w_{i+1/2}-w_{i-1/2}}{x_{i+1/2}-x_{i-1/2}} ,\\
w|_{x=x_{i+1/2}} \approx &&\mathfrak{I}_{i+1/2}(w_{i+1},w_i) = \frac{1}{2}(w_{i+1}+w_i),\\
\label{op4} w|_{x=x_i} \approx &&\mathfrak{I}_{i}(w_{i+1/2},w_{i-1/2}) = \frac{w_{i+1/2}(x_i-x_{i-1})+w_{i-1/2}(x_{i+1}-x_i)}{x_{i+1}-x_{i-1}}.
\end{eqnarray}
A standard error analysis indicates that each operator is second order, \emph{i.e.}, that the errors are expected to decrease proportional to the grid spacing squared.

The discrete operators defined by (\ref{op1})-(\ref{op4}) can be used to approximate the electric field, and electron and hole currents at the half-points. The discretised versions of equations (\ref{phieqn}b), (\ref{neqn}b) and (\ref{jp}b) are 
\begin{eqnarray}
E_{i+1/2} = -\mathfrak{D}_{i+1/2} \left( \phi_{i+1},\phi_i \right),\\
\jn_{i+1/2} = \kappa_n \left( \mathfrak{D}_{i+1/2}(n_{i+1},n_i) + \mathfrak{I}_{i+1/2}(n_{i+1},n_i) E_{i+1/2} \right),\\
\jp_{i+1/2} = -\kappa_p \left( \mathfrak{D}_{i+1/2}(p_{i+1},p_i) - \mathfrak{I}_{i+1/2}(p_{i+1},p_i) E_{i+1/2}\right),\\
\text{for} \quad i=0,...,N-1.\nonumber
\end{eqnarray}
The anion vacancy flux can then be computed on both the internal and boundary grid points by discretising equation (\ref{Peqn}b) and its boundary conditions (\ref{leftbcs}a) and (\ref{rightbcs}a). We have
\begin{eqnarray}
\fp_1 = \fp_N = 0,\\ 
\fp_i = -\mathfrak{D}_i (P_{i+1/2},P_{i-1/2}) + \mathfrak{I}_i (P_{i+1/2},P_{i-1/2}) \mathfrak{D}_i (E_{i+1/2},E_{i-1/2}),\\
\text{for} \, \, i=1,..,N-1.\nonumber
\end{eqnarray}
The ODEs governing the evolution of the anion vacancy density, arising from (\ref{Peqn}a), are
\begin{eqnarray}
\label{Pdisc}\frac{dP_{i+1/2}}{dt} = -\lambda \mathfrak{D}_{i+1/2} ( \fp_{i+1},\fp_{i} ),\\
\text{for} \quad i=0,...,N-1.\nonumber
\end{eqnarray}
The algebraic equations for the potential resulting from discretisation of (\ref{phieqn}a) and the boundary conditions (\ref{leftbcs}b) and (\ref{rightbcs}b) are
\begin{eqnarray}
\label{phidisc1} 0 = \phi_0 - \frac{\Phi-\Phi_{bi}}{2},\\
0 = \lambda^2 \mathfrak{D}_i (E_{i+1/2},E_{i-1/2}) + (1-\mathfrak{I}_i (P_{i+1/2},P_{i-1/2}) + \delta (n_i - p_i )),\\
\text{for} \, \, i=1,..,N-1,\nonumber \\*[2mm]
\label{phidisc3} 0 = \phi_N + \frac{\Phi-\Phi_{bi}}{2}.
\end{eqnarray}
It is straightforward to discretise (\ref{neqn}a) and (\ref{jp}a) to formulate ODEs governing the evolution of the electron and hole densities on the internal grid points. For the electrons (holes) a Dirichlet-type boundary condition is to be imposed on $x=0$ ($x=1$), see (\ref{leftbcs}c) ((\ref{rightbcs}c)), whilst a specified value of the flux is to be imposed at the boundary $x=1$ ($x=0$), see (\ref{rightbcs}d) ((\ref{leftbcs}d)). The former of these conditions immediately specifies $n_0=\nt$ ($p_N=\pt$) whilst the latter is imposed by linear extrapolation of the electron (hole) current to the boundary $x=0$ ($x=1$). Hence we arrive at the following system of ODEs and algebraic equations which are sufficient to determine the values of electron density at all grid points.
\begin{eqnarray}
\label{ndisc1} 0 = n_0 - \nt,\\
\nu \frac{dn_i}{dt} = \mathfrak{D}_{i} (\jn_{i+1/2},\jn_{i-1/2}) + G(x_i) + R(n_i,p_i),\\
\text{for} \, \, i=1,..,N-1,\nonumber \\*[2mm]
\label{ndisc3} 0 = \jn_{N-1/2} + \mathfrak{D}_{N-1}(\jn_{N-1/2},\jn_{N-3/2}) (1-x_{N-1/2}) + R_r(n_N).
\end{eqnarray}
Similarly, the equations determining the hole density at each grid point are
\begin{eqnarray}
\label{pdisc1} 0 = \jp_{1/2}-\mathfrak{D}_{1}(\jp_{3/2},\jp_{1/2}) x_{1/2} + R_l(p_0),\\
\nu \frac{dp_i}{dt} = -\mathfrak{D}_{i} (\jp_{i+1/2},\jp_{i-1/2}) + G(x_i) + R(n_i,p_i),\\
\text{for} \, \, i=1,..,N-1,\nonumber \\*[2mm]
\label{pdisc3} 0 = p_N-\pt.
\end{eqnarray}
We note that the boundary conditions (\ref{ndisc3}) and (\ref{pdisc1}) are the only conditions not imposed exactly. However, the extrapolation involved only introduces second order errors, and therefore the scheme as a whole is still second order as demonstrated by figure \ref{convplot}.

\section{\label{implementation}Implementation}
\begin{figure}
\centering
\includegraphics[width=0.4\textwidth]{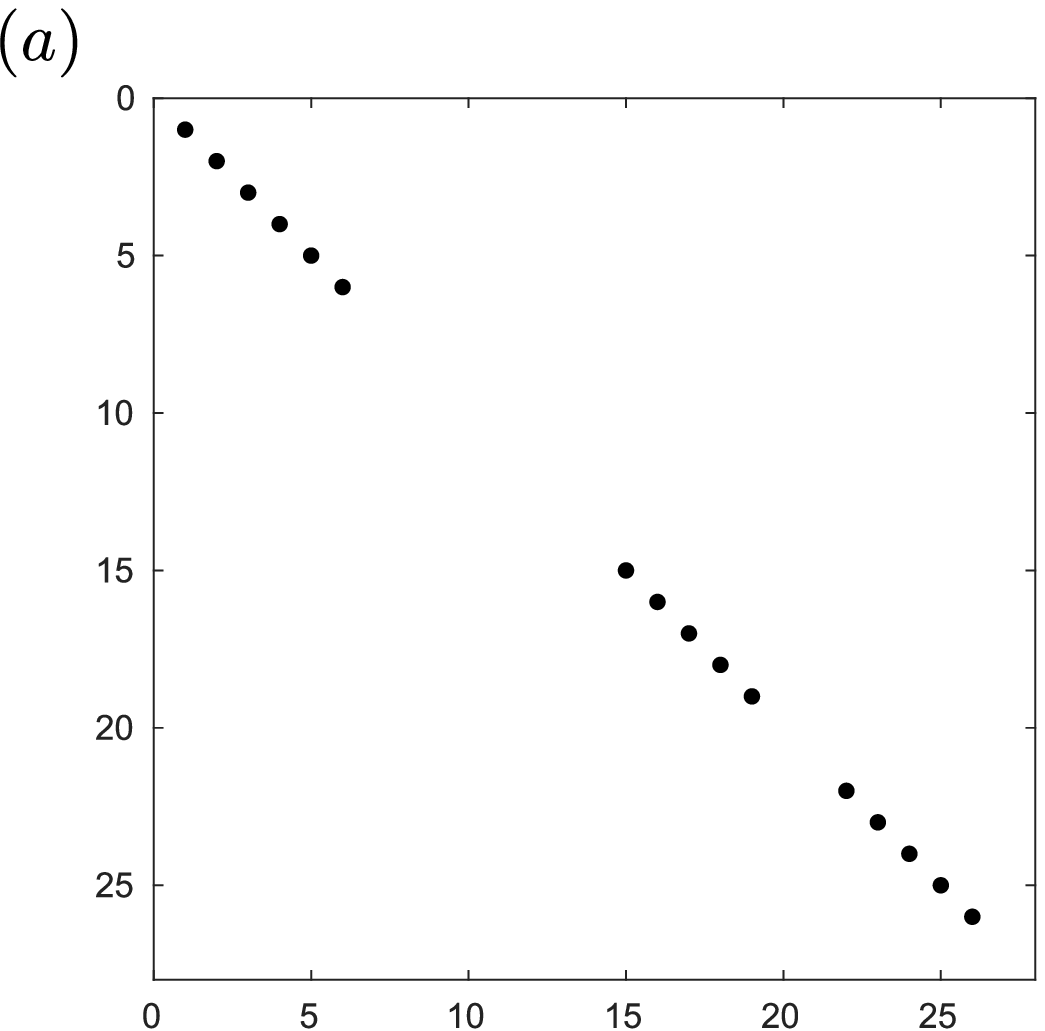} \hspace{1em}
\includegraphics[width=0.4\textwidth]{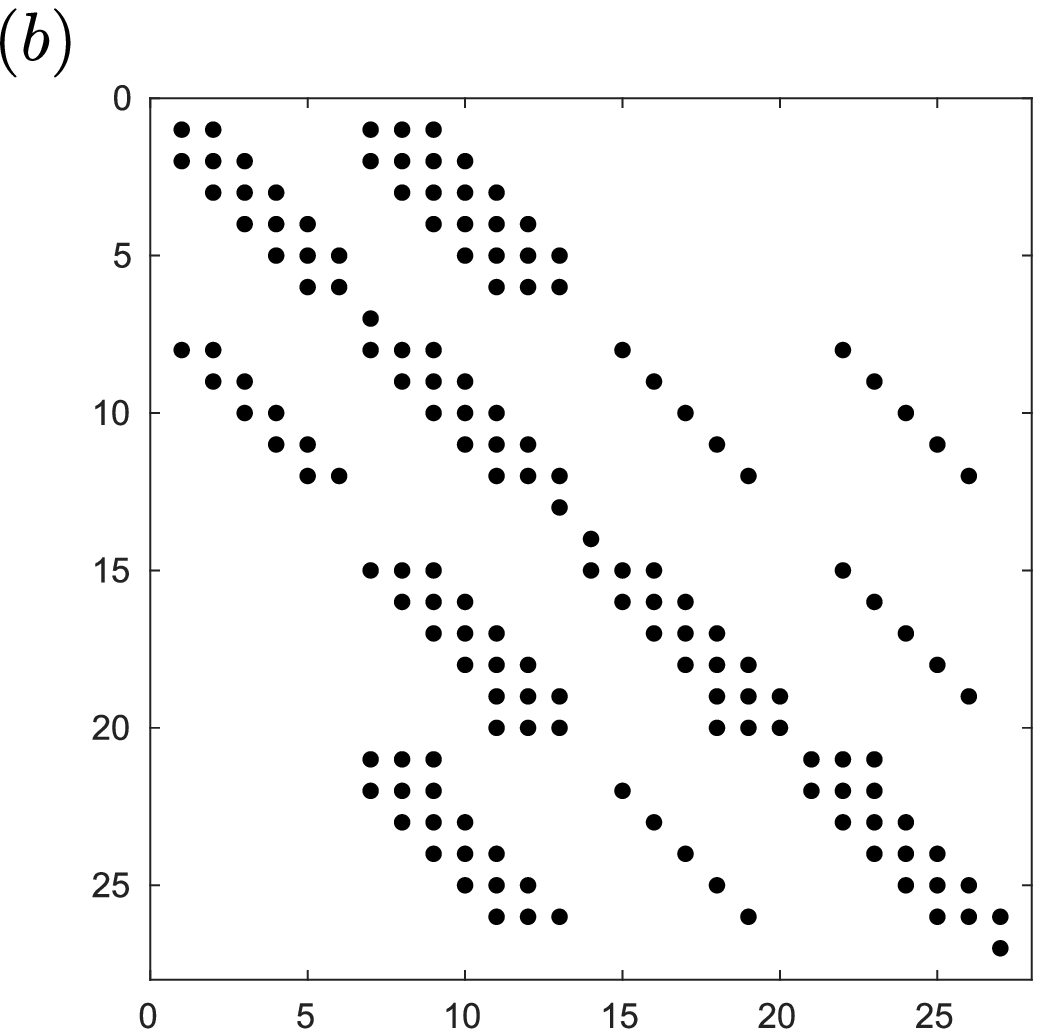}
\caption{Plots to show the positions of the nonzero entries (black dots) in the mass matrix $\MM$, panel (a), and the Jacobian matrix of $\ff$, panel (b), in the case $N=6$.}
\label{fig11}
\end{figure}

The system of DAEs formulated above is evolved forward in time using MATLAB's {\tt ode15s}. In order to minimise computational cost, by minimising the size of the system, we eliminate superfluous variables (namely $E_{i+1/2}$, $\jp_{i+1/2}$, $\jn_{i+1/2}$ and $\fp_i$) and assemble the remaining $4N+3$ unknown functions of time into one column vector $\uu(t)$ as follows:
\be 
\uu(t) = \left[ P_{1/2}, \cdots, P_{N-1/2}, \phi_0, \cdots, \phi_N, n_0, \cdots, n_N, p_1, \cdots ,p_N \right]^{\text{T}},\\
\label{vec} \mbox{or equivalently} \quad \uu(t) = \left[ \PP(t)^{\text{T}} \, \,  \bphi(t)^{\text{T}} \, \, \nn(t)^{\text{T}} \, \, \pp(t)^{\text{T}} \right] ^{\text{T}},
\ee
where a superscript T denotes a transpose. In (\ref{vec}) $\PP$ is a column vector of length $N$ whose entries are the functions of time $P_{i+1/2}(t)$ for $i=0,...,N-1$. Similarly, $\bphi$, $\nn$ and $\pp$ are column vectors of length $N+1$ whose entries are the functions $\phi_i(t)$, $n_i(t)$ and $p_i(t)$ respectively. The problem to be solved can now be written in the form
\begin{equation}
\MM \frac{d\uu}{dt} = \ff(\uu) \quad \text{with} \quad \uu|_{t=0}=\uu_0, 
\end{equation}
where $\ff(\uu)$ is a nonlinear vector function of length $4N+3$ whose first $N$ entries are the right-hand sides of (\ref{Pdisc}), the subsequent $N+1$ entries are the right-hand sides of (\ref{phidisc1})-(\ref{phidisc3}), followed by the $N+1$ right-hand sides of (\ref{ndisc1})-(\ref{ndisc3}), and finally the $N+1$ right-hand sides of (\ref{pdisc1})-(\ref{pdisc3}). The matrix $\MM$ is a $(4N+3)\times(4N+3)$ diagonal mass matrix whose entries are the coefficients of the time derivative terms in the equations (\ref{Pdisc})-(\ref{pdisc3}), see figure \ref{fig11}. Since the governing equations for the values of $\phi_i$, (\ref{phidisc1})-(\ref{phidisc3}), and the discrete boundary conditions (\ref{ndisc1}), (\ref{ndisc3}), (\ref{pdisc1}) and (\ref{pdisc3}) contain no temporal derivatives the corresponding entries on the diagonal of $\MM$ are zero and hence the mass matrix is singular. It is this feature of the system that renders the problem a system of DAEs and motivates our choice of solver, namely {\tt ode15s}, which is one of relatively few solvers that can handle problems of this type. Moreover, it offers an adaptive timestep which is able to deal with the disparity in timescales between the electronic and ionic motion both with good accuracy and without high computational penalties.

The {\tt ode15s} time step requires numerical approximation of the Jacobian of $\ff$. However, it is clear from the structure of the discrete system that many entries in the Jacobian are zero. A heavy reduction in computational effort, around $0.005 N^{3/2}$ in the convergence test case, can be achieved by exploiting the {\tt jpattern} option which allows the user to `flag' only a subset of the entries in the Jacobian matrix that need to be numerically approximated (those that are not flagged are assumed to be zero) at each time step, see figure \ref{fig11}.

When simulating many of the regimes of physical interest, we find that the condition number of the Jacobian is large, sometimes $O(10^{16})$ or greater, which has the potential to severely hamper the accuracy of matrix inversions performed by the solver, {\tt ode15s}. To overcome this difficulty we make use of a third-party toolbox, Advanpix, which extends MATLAB's functionality to work with floating point numbers of higher precision. This can result in a loss in the speed of computation for simple cases (those which can be computed accurately on a small number of grid points) for which runtimes, originally of a few seconds on a standard desktop computer, are typically 15-30 times longer. However, the higher precision offered by Advanpix's Multiprecision Computing Toolbox \cite{mct17} is crucial in allowing the numerical scheme to cope with the \emph{stiff} regimes of practical interest.

\section{\label{verify}Verification and test cases}
We verify the performance of the scheme using  physically relevant test cases. Here we choose to compare our numerical results to the asymptotic solution given in \cite{Courtier2017} in the particular case where the bulk recombination, $R(n,p)$, is monomolecular (depending solely on the local hole concentration $p$) and the surface recombination rates, $R_l(p)$ and $R_r(n)$ are both zero, such that
\be
R(n,p) = \gamma p, \quad R_l(p) = R_r(n) = 0.
\ee
Here $\gamma$ is a dimensionless rate constant. As noted in \cite{Courtier2017}, monomolecular hole-dominated bulk recombination is the limit of the Shockley-Read-Hall (SRH) recombination law when electron pseudo-lifetime is much less than hole pseudo-lieftime. This is a reasonable (though not completely accurate) description of recombination in the perovskite material methylammonium lead tri-iodide \cite{Stranks2014}. Note however that the numerical scheme presented here is readily able to deal with nonlinear recombination rates such as the full SRH law. As is typical, in such applications, we assume that the photo-generation rate, $G(x)$, obeys the Beer-Lambert law, which has the dimensionless form 
\begin{eqnarray}
G(x) = \Upsilon \exp(-\Upsilon x).
\end{eqnarray} 
in which $1/\Upsilon$ is the dimensionless absorption length. Estimates of these from \cite{Courtier2017} are
\be
\Upsilon \approx 3.7, \qquad \gamma \approx 2.4.
\ee
A simple choice of initial conditions satisfying the boundary conditions is 
\begin{eqnarray}
P_{\text{init}} = 1, \quad n_{\text{init}} = \nt, \quad p_{\text{init}} = \pt.
\end{eqnarray}
In practice, the choice of initial conditions is not particularly important for the modelling of PSCs as, in any experimental procedure, it is standard practice to include a pre-conditioning step. This step involves holding the applied potential $\Phi(t)$ constant for a sufficiently long time such that any initial transients associated with charge carrier and ion vacancy motion decay towards zero.

\subsection{Choice of spatial grid}
In anticipation of the fact that the largest gradients in the solution appear in narrow Debye layers adjacent to the domain boundaries, we use a spatial grid comprised of Chebyshev nodes on the interval $[0,1]$ defined as follows
\be
x_i = \frac{1}{2} \left( 1+ \cos \left[ \pi \left( \frac{i}{N} - 1 \right) \right] \right), \quad \mbox{for} \quad i=0,...,N.
\ee
By concentrating points near the domain boundaries we are able to achieve good resolution in the Debye layers without wasting computational effort by over resolving in the bulk where the solution varies more slowly.

\subsection{Spatial convergence} 
Here we verify the expected second order pointwise convergence of the scheme by testing the method in a scenario in which an illuminated cell is initially operating at an applied voltage equal to the built-in voltage $\Phi=\Phi_{bi}=40$ (see \ref{stack}), so that the potential drop across the perovskite layer is zero. The applied voltage is then rapidly decreased to $\Phi=20$ so that the device is running in its power generating regime. We accomplish this by defining
\begin{equation}
\Phi(t) = \Phi_{bi} - \frac{20 \tanh(\beta t)}{\tanh(\beta t_{\text{end}})} \quad \mbox{with} \quad \beta = 10^2, \quad t_{end}=1.
\end{equation}
Here $\beta$ characterises the timescale for altering the applied voltage and $t_{\text{end}}$ is the time at which the simulation is terminated. We terminate the simulation at $t=t_{end}$ which is not so large that the solution has reached steady state. During this convergence test we are therefore examining the scheme's accuracy during the transient and not just at steady state.

To demonstrate the pointwise convergence of the scheme, we monitor the values of five representative quantities, namely
\be \label{measures}
n|_{x=0,t=t_{end}}, \quad n|_{x=1/2,t=t_{end}}, \quad p|_{x=1/2,t=t_{end}}, \quad p|_{x=1,t=t_{end}}, \quad \phi|_{x=1/2,t=t_{end}}.
\ee
These are chosen because, independent of the number of grid points used, they are available immediately following the temporal integration without the need for an additional interpolation step, thereby allowing a direct interrogation of the scheme.

\begin{figure}
\centering
\includegraphics[width=0.65\textwidth]{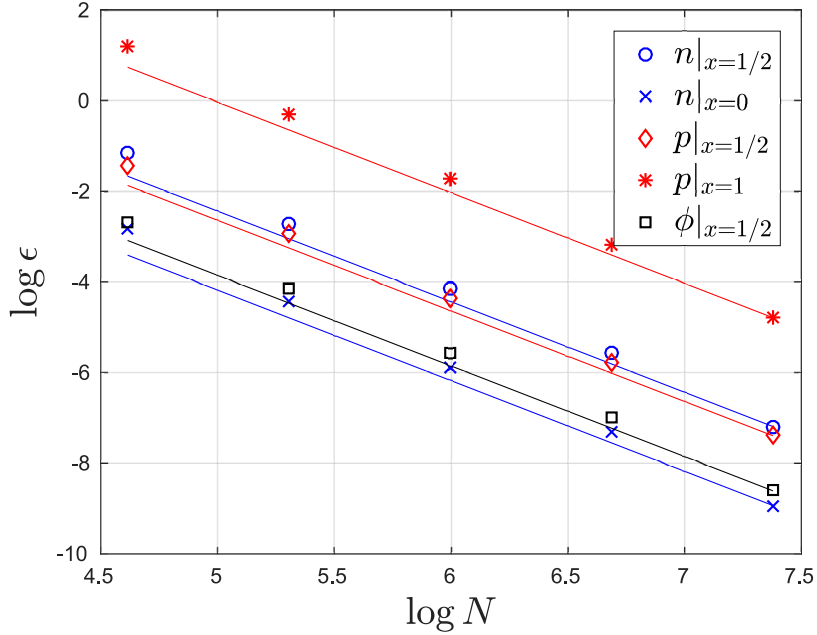}
\caption{A plot showing the pointwise convergence of the scheme with plots of the pointwise error, $\epsilon$, versus the number of grid points, $N$. Markers show errors for computations with $N=100,200,400,800,1600$ estimated against a numerical solution with $N_M=3200$ grid points. Lines show (the expected) second order convergence.}
\label{convplot}
\end{figure}

Due to the lack of exact solutions to the model, it is necessary to verify convergence by measuring the error relative to a solution computed on a highly refined grid. For some scalar quantity $v$ (which could be any of those defined in (\ref{measures})) computed using the scheme on $N$ grid points, denoted by $v^{(N)}$, we can define the error,
\be
\epsilon (N) = | v^{(N)} - v^{(N_M)} |,
\ee
for some $N_M \gg N$. The pointwise convergence of the quantities in (\ref{measures}) is shown in figure \ref{convplot}. These results which demonstrate second order order convergence of the scheme, as predicted by the naive analysis of (\ref{op1})-(\ref{op4}).

\subsection{A current decay transient: comparison between numerical and asymptotic solutions}

\begin{figure}
\centering
\includegraphics[width=0.6\textwidth]{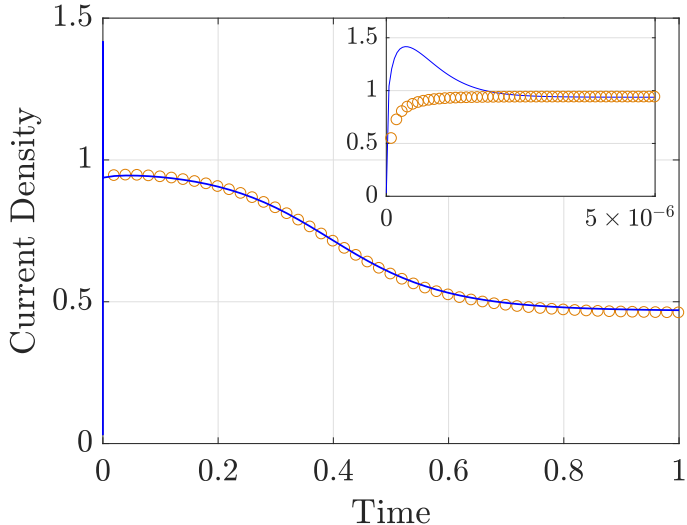}
\caption{Current as a function of time. Inset shows the fast initial electronic transient not captured by the asymptotics. Blue solid lines represent the full numerical solutions and orange circles represent the uniformly-valid asymptotic expansions.}
\label{JumpCurrent}
\end{figure}

\begin{figure}
\centering
\includegraphics[width=0.45\textwidth]{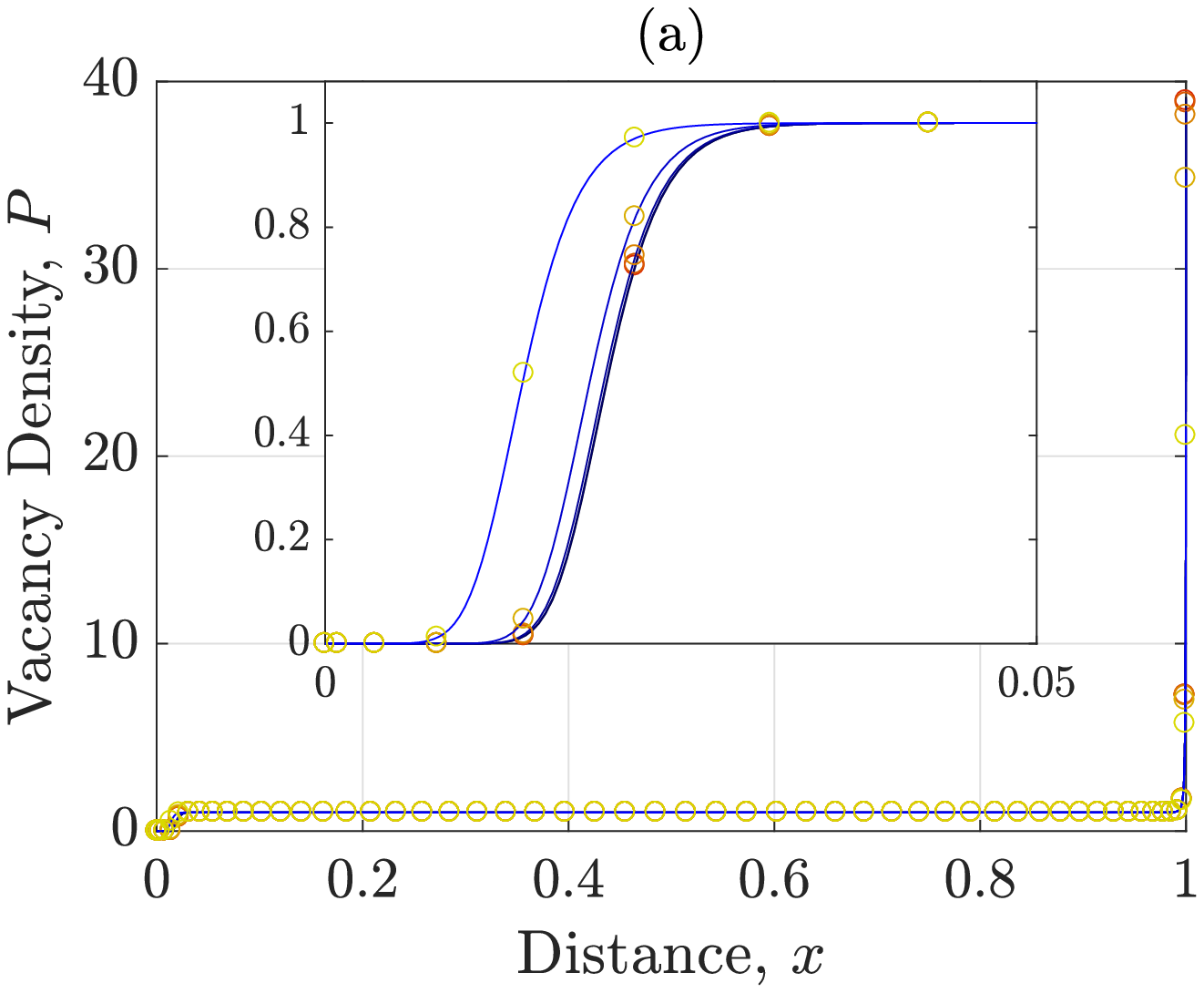}
\includegraphics[width=0.45\textwidth]{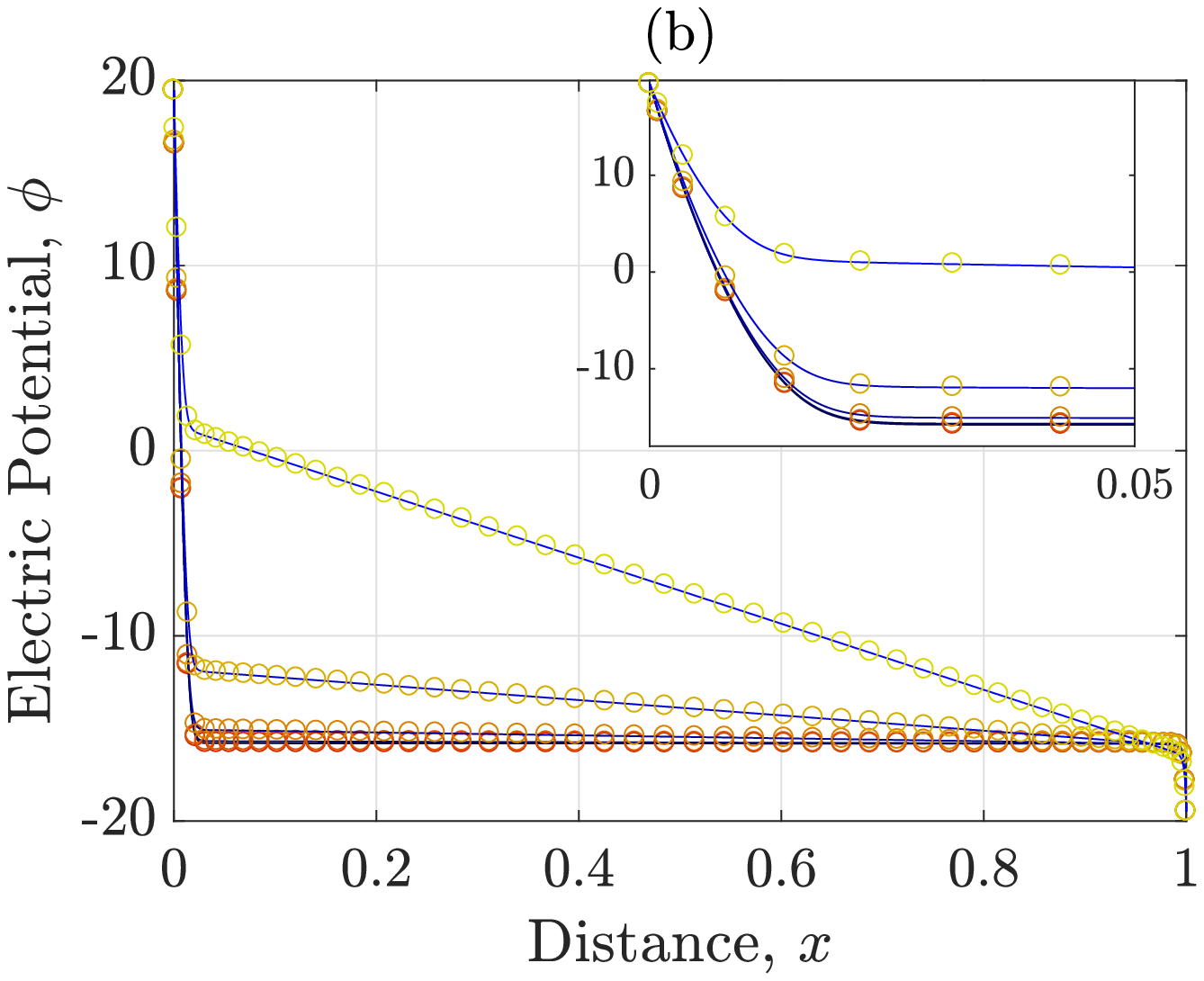} \\
\includegraphics[width=0.45\textwidth]{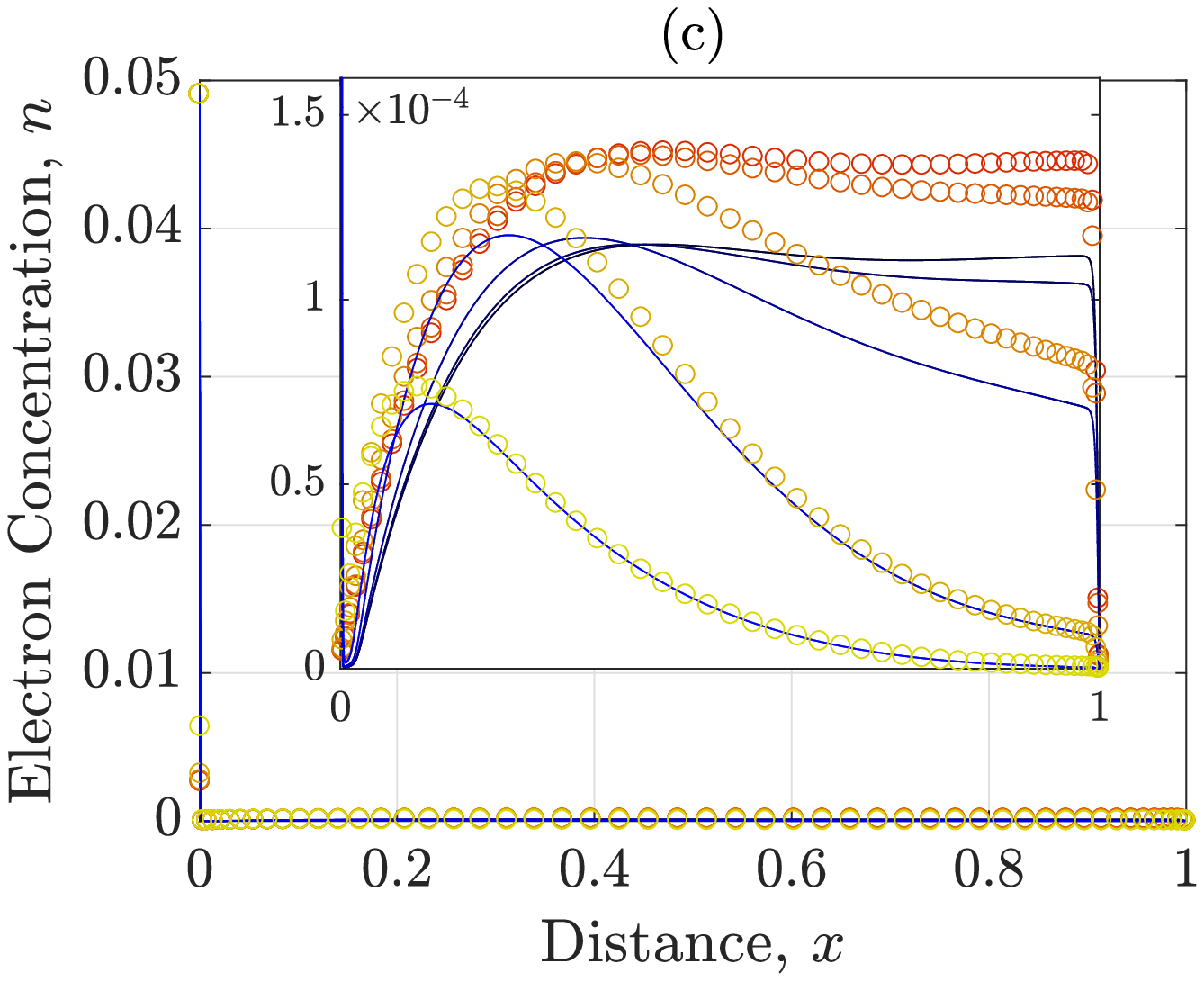}
\includegraphics[width=0.45\textwidth]{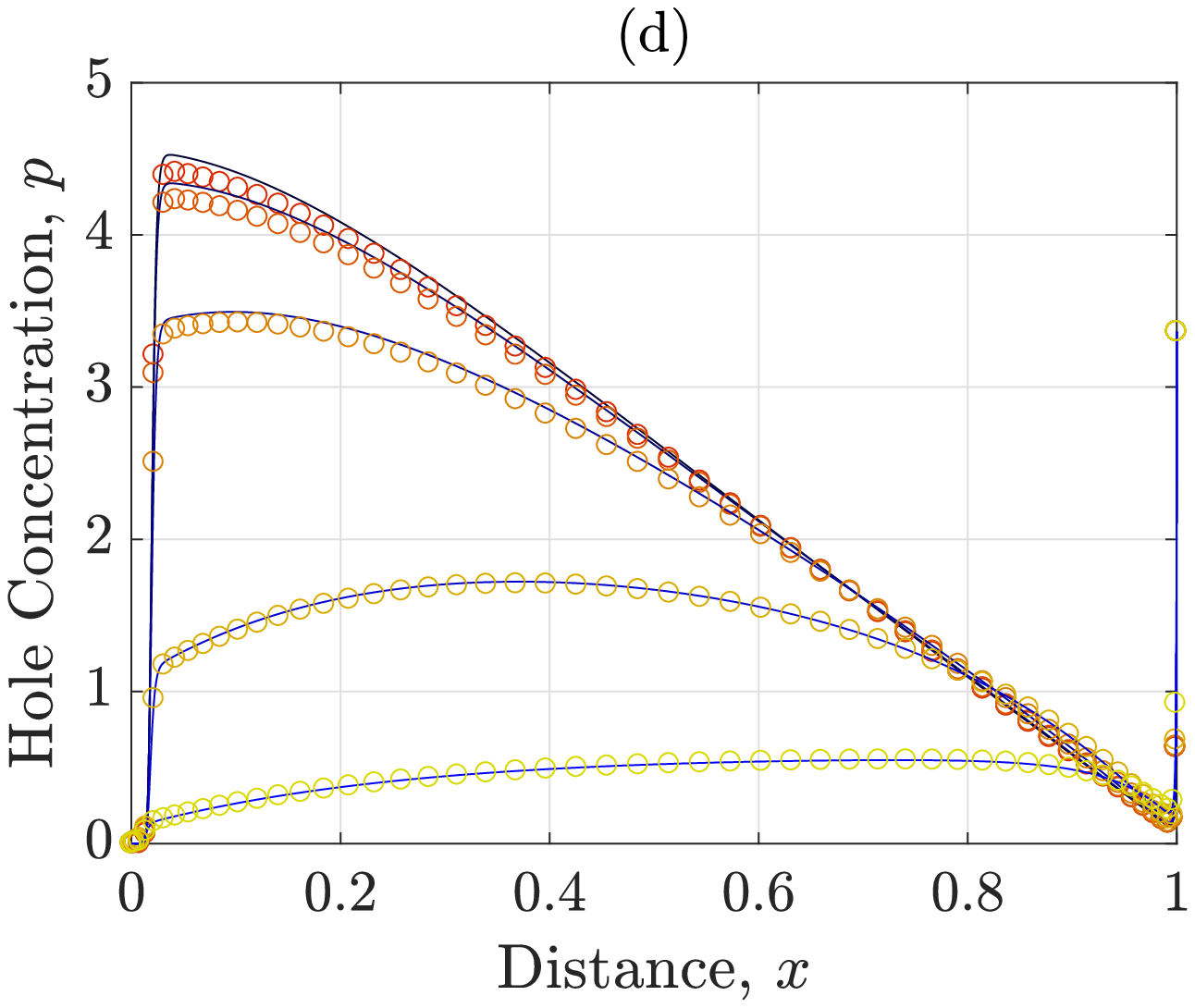}
\caption{(a) Anion vacancy density, (b) electric potential, (c) electron concentration and (d) hole concentration profiles across the perovskite layer of a PSC at $t=0.2,0.4,0.6,0.8,1$. Solid lines (blue to black for increasing time) represent the full numerical solutions and circles (yellow to red for increasing time) represent the uniformly-valid asymptotic expansions.}
\label{JumpProfiles}
\end{figure}

Here we simulate a cell that is preconditioned at $\Phi=\Phi_{bi}$ for a sufficient time to eliminate transients, before undergoing a smooth but rapid decrease in applied bias from $\Phi=\Phi_{bi}$ at $t=0$ to $\Phi=0$ some short time later, obeying,
\begin{equation}
\Phi(t) = \Phi_{bi}\left(1 - \frac{\tanh(\beta t)}{\tanh(\beta t_{\text{end}})}\right) \quad \mbox{with} \quad \beta = 10^6, \quad t_{end}=1.
\end{equation}
A comparison between the photocurrent calculated from the numerical solution (as described above) and the asymptotic solution (as described in \cite{Courtier2017}) is made in figure \ref{JumpCurrent}, showing remarkable agreement between the methods for all but very small times. The inset shows the very short time behaviour where there is significant discrepancy between the two solutions. This results from the numerical method capturing fast electronic transients (associated with electron and hole motion) that are absent from the asymptotic solution. The ability to simulate on the timescale of electronic transport through to that of ion motion is important for rigorous investigation of many properties of PSCs.

In figure \ref{JumpProfiles}, we demonstrate good agreement between the numerical and asymptotic solutions, with the noticeable exception of panel (c) for the electron distribution in which the solutions vary in magnitude, but not shape. From a convergence check (increasing both spatial and temporal accuracy), it can be confirmed that the numerical scheme has indeed converged. The discrepancy arises from a small $O(10^{-2})$ error in the asymptotic solution to the electric potential which in turn leads to the significant error in the asymptotic solution for electron density $n$, that can be seen panel (c).

\subsection{A current-voltage curve: comparison between numerical and asymptotic solutions}

\begin{figure}
\centering
\includegraphics[width=0.6\textwidth]{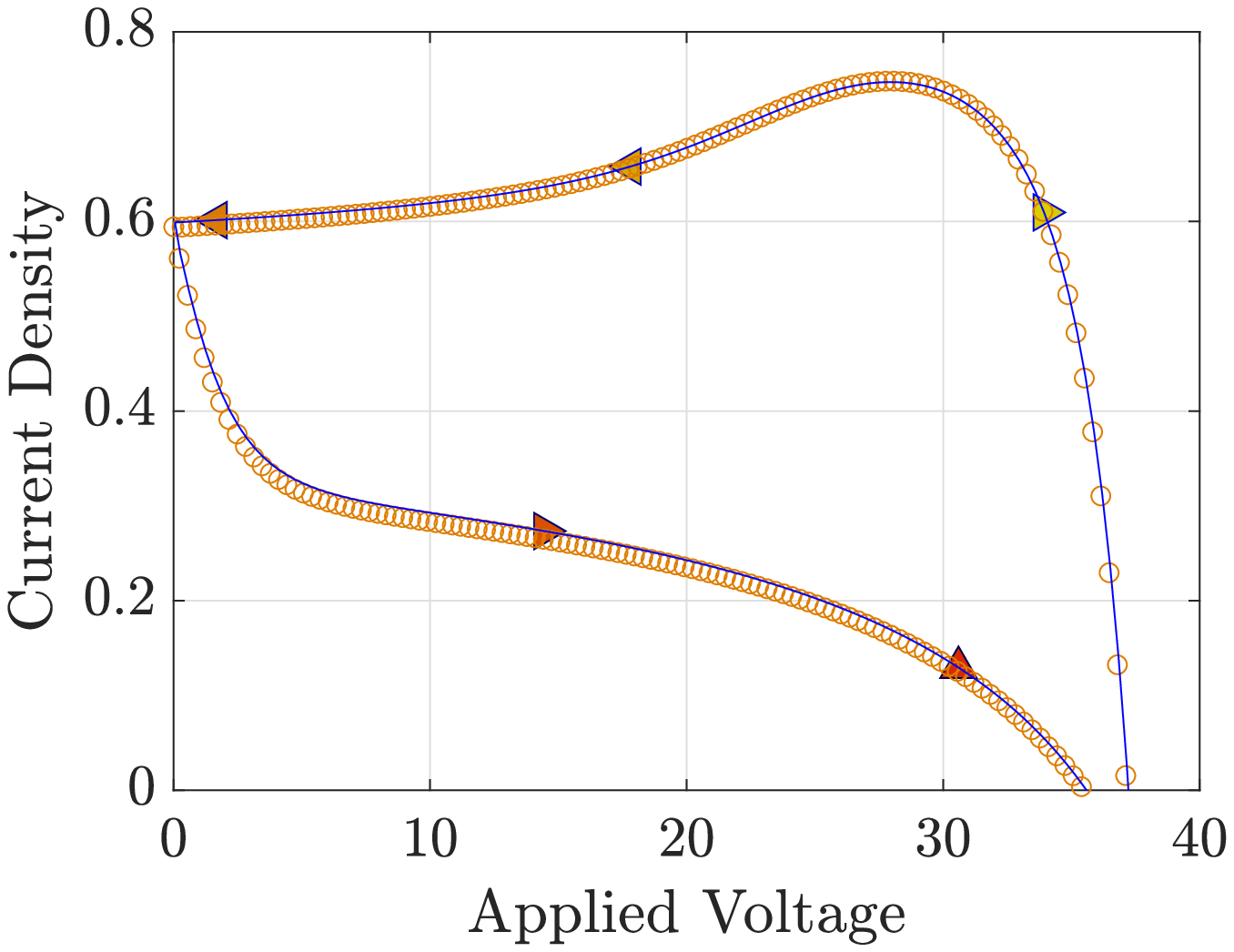}
\caption{Current as a function of applied voltage during a J-V scan. The blue line represents the full numerical solution, orange circles represent the uniformly-valid asymptotic expansions and the triangles (yellow to red for increasing time) indicate times corresponding to plots in figure \ref{JVProfiles}.}
\label{JVCurve}
\end{figure}

\begin{figure}
\centering
\includegraphics[width=0.45\textwidth]{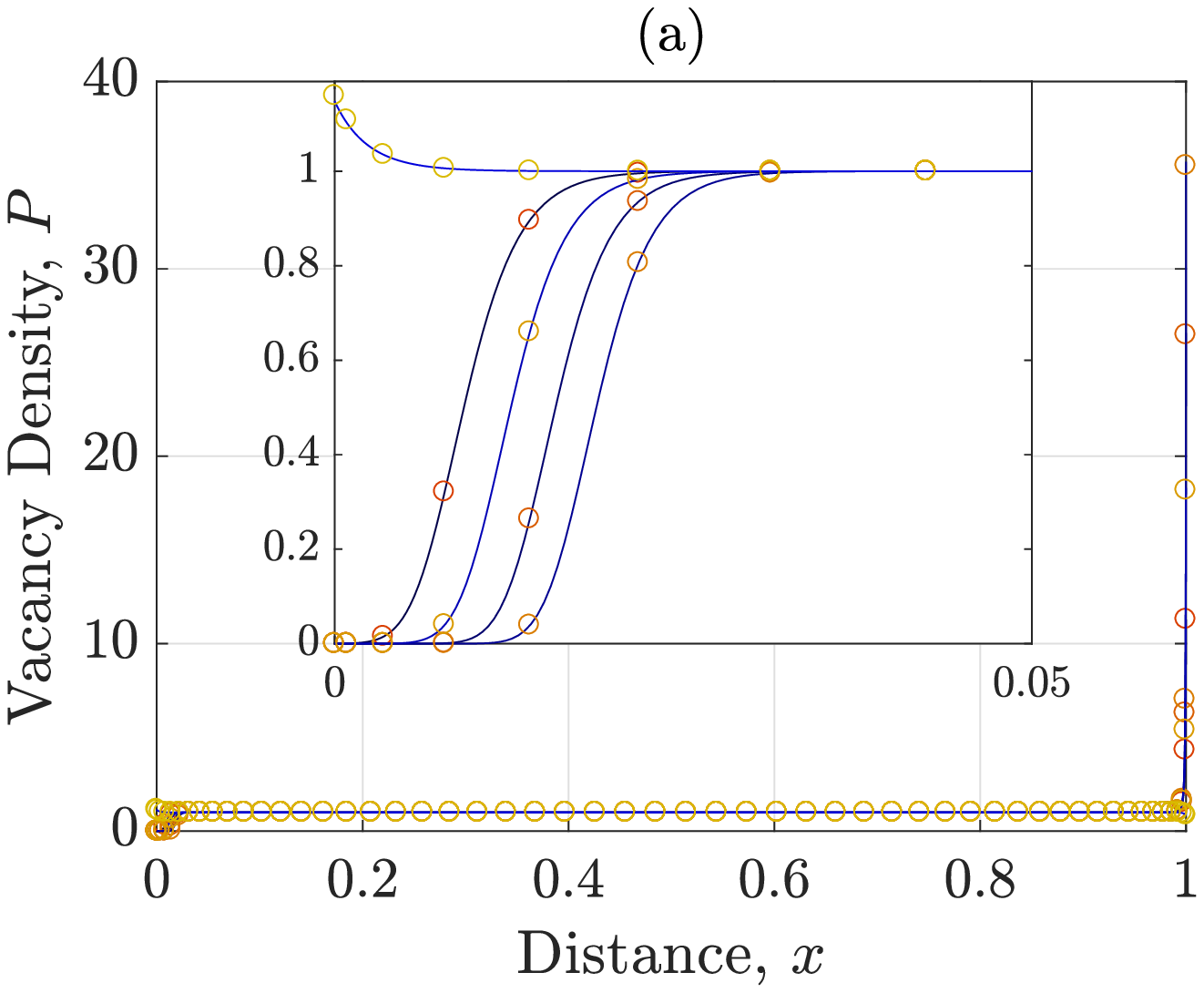}
\includegraphics[width=0.45\textwidth]{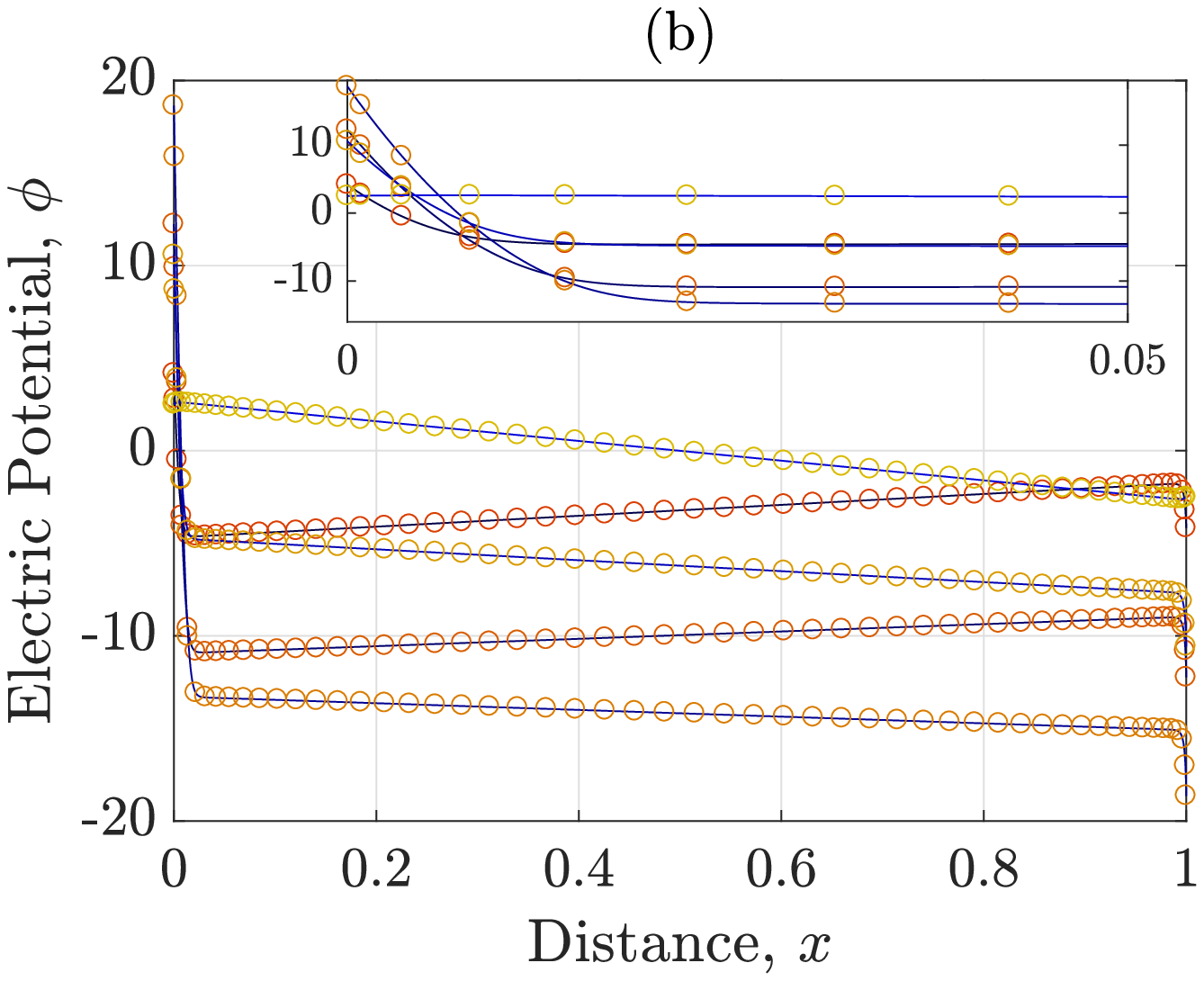} \\
\includegraphics[width=0.45\textwidth]{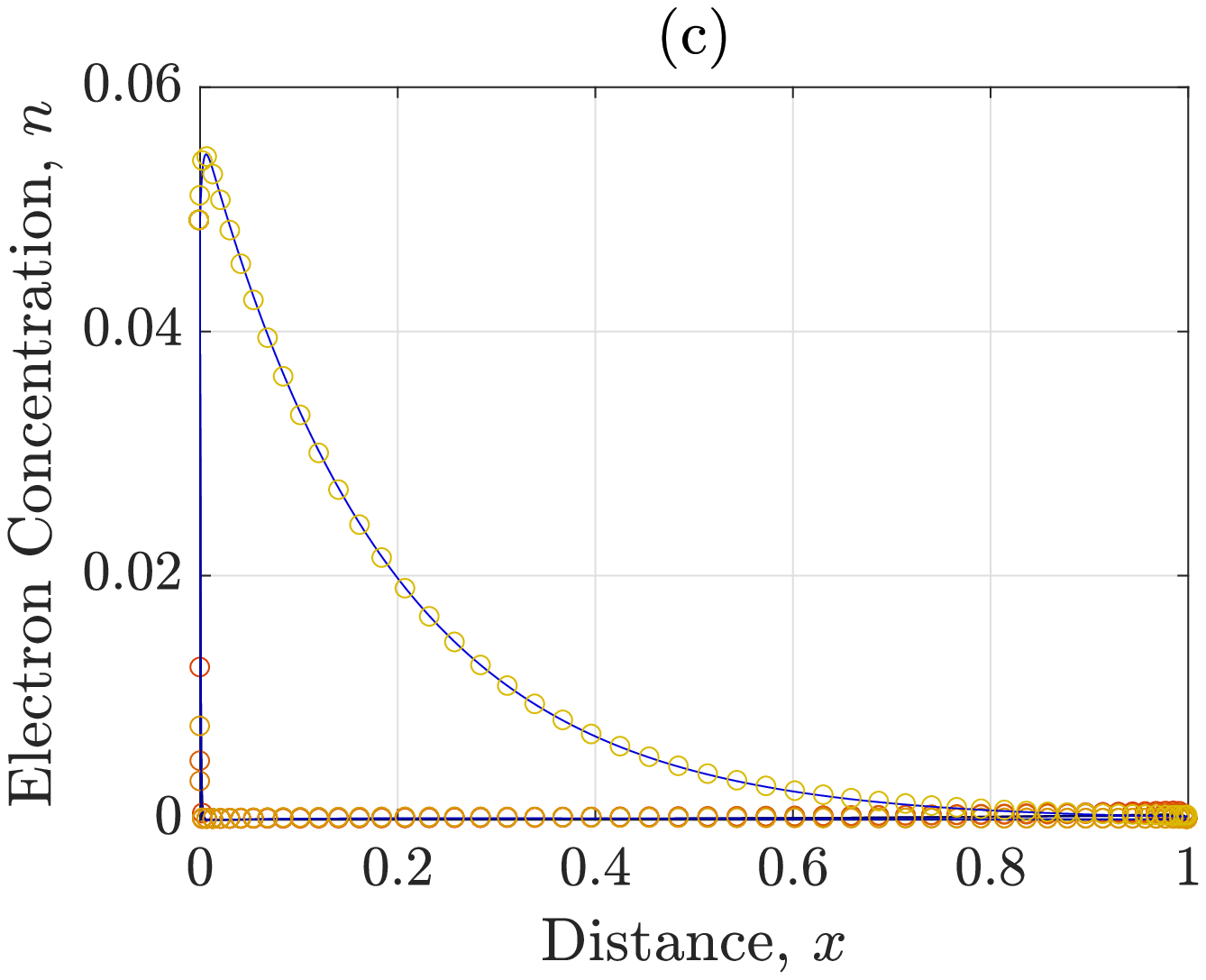}
\includegraphics[width=0.45\textwidth]{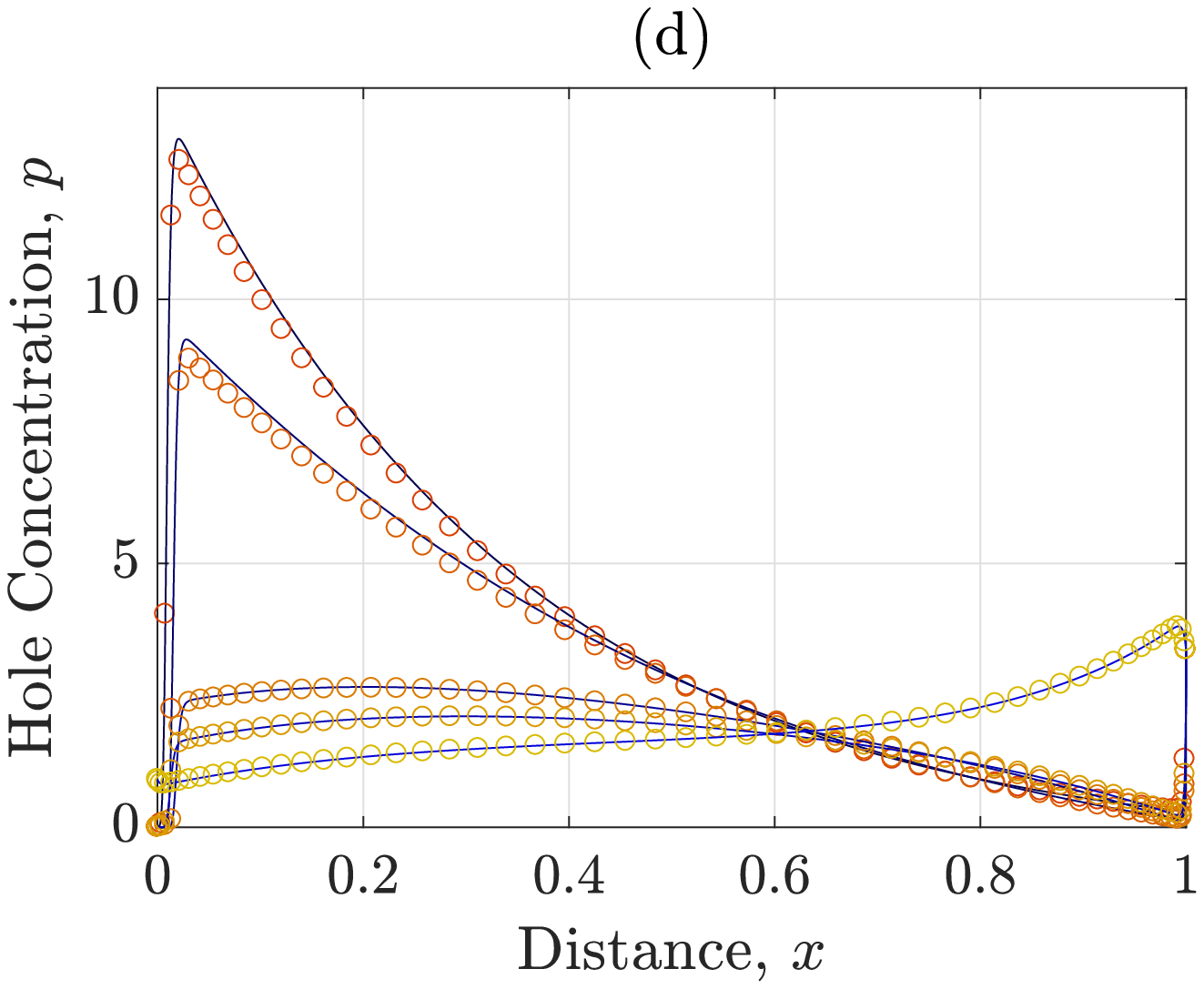}
\caption{As for figure \ref{JumpCurrent}, with plots at 5 equally spaced times during a JV-scan.}
\label{JVProfiles}
\end{figure}

Measurements of the current generated by a solar cell in response to a backwards and forwards sweep of the applied voltage are a common way of measuring cell properties and, in the case of PSCs, frequently result in significant sweep-rate dependent `hysteresis' \cite{Snaith2014}. This `hysteresis' is a signature of the slow timescale ion motion in the cell.  The standard experimental procedure for generating such a $J$-$V$ hysteresis-curve, including a preconditioning step, will be simulated as follows. The cell is preconditioned in forward bias (at 1.2V, corresponding to $\Phi\approx 46.7$), then the voltage is scanned smoothly from forward bias ($\Phi_{ap}>\Phi_{bi}$)  to short circuit ($\Phi_{ap}=0$) and back at a scan rate equivalent to 100 mV/s (corresponding to a rate of $\approx 7.1$ in dimensionless units). The resulting J-V curve is compared to the corresponding asymptotic solution from \cite{Courtier2017} in figure \ref{JVCurve}. Plots of the solutions variables in space, at equally-spaced times through the hysteresis sweep, are made in figure \ref{JVProfiles}.

\section{\label{discussion}Discussion}
We previously attempted to solve this \emph{stiff} numerical problem using spectral methods via the Chebfun package \cite{chebfun} for MATLAB. An advantage of spectral methods is the high spatial accuracy that can be obtained with a small number of grid points. However, this method requires such fine resolution in time to maintain accuracy that it becomes too computationally expensive to solve problems in the desired parameter regimes. In particular, the ratio of the Debye length to the perovskite layer width, $\lambda$, caused a significant slowdown if taken to be $O(10^{-2})$ or smaller, and as a result solutions for the realistic value of $\lambda=2.4\times 10^{-3}$ were unobtainable in a reasonable amount of time (days).

Calado \etal\cite{Calado2016} use MATLAB's built-in solver PDEPE to investigate a drift-diffusion model for PSCs. However, we found that in realistic parameter regimes this method has a condition number that is so large that matrix inversions cannot be performed to within even single-digit accuracy. The method could thus only be applied to {\it toy} problems in order to give qualitative indications of the physical phenomena occurring in real cells. The increases in accuracy afforded by the staggered grid and the increased arithmetic precision used in our approach are essential in order to investigate the behaviour of real PSCs.
A secondary disadvantage of using PDEPE is that it can only accept certain types of boundary conditions and this prevents the extension of the method to a three-layer model that incorporates charge carrier transport in the ETL and the HTL of a PSC. In particular PDEPE is unable to deal with conditions posed on internal boundaries such as occur at the ETL/perovskite and the perovskite/HTL interfaces. In contrast, the tailored numerical scheme detailed in this work can be readily extended to include more layers and incorporate any necessary type of interface condition.

\section{\label{conclusions}Conclusions}
We have presented a numerical scheme for solving a PDE drift-diffusion model of ion vacancy and charge carrier transport in a perovskite solar cell. The scheme uses second-order finite difference approximations of spatial derivatives, on a non-uniform staggered mesh, to reduce the system of PDEs to a (large) system of ordinary differential-algebraic equations (with time as the independent variable). This system of ordinary differential-algebraic equations is solved with the MATLAB routine {\tt ode15s} in conjunction with Advanpix's Multiprecision Computing Toolbox \cite{MATLAB,mct17}. 

For realistic parameter values and in appropriate operating regimes, the problem exhibits significant stiffness owing to (i) the small Debye length, (ii) large potential differences across the device (giving rise to large and rapid changes in solution across narrow Debye layers), and (iii) vastly different timescales for the transport of ion vacancies and electronic charge carriers. We were able to circumvent these difficulties by (a) using a non-uniform mesh to selectively concentrate grids points in the Debye layers in order to obtain high accuracy without prohibitive computational cost, and (b) using an adaptive timestep. In doing so we have presented the only published method for numerical solution of a truly realistic charge transport model for a metal halide perovskite solar cell. Moreover, the method can be used to simulate experimental protocols, such as current decay transients and current-voltage sweeps, in only a few minutes on a desktop computer.

\paragraph{Acknowledgements} NEC is supported by an EPSRC funded studentship from the CDT in New and Sustainable Photovoltaics. All authors are also grateful to B. Protas, A. B. Walker, J. Cave and S. E. J. O'Kane for numerous insightful discussions that led to substantial improvements in the work.

\bibliographystyle{siam}
\bibliography{NumericsBib}

\end{document}